\documentclass{aastex}

\usepackage{graphicx} 
\usepackage[english]{babel}
\usepackage{hyperref}

\begin{document}

\title{Asymptotic Steady State Solution to a Bow Shock with an Infinite Mach Number}

\author{A. Yalinewich\altaffilmark{1} and R. Sari\altaffilmark{1,2}}
\altaffiltext{1}{Racah Institute of Physics, the Hebrew University, 91904, Jerusalem, Israel}
\altaffiltext{2}{California Institute of Technology, MC 350-17, Pasadena, CA 91125}
\date{\today}

\begin{abstract}
The problem of a cold gas flowing past a stationary object is considered. It is shown that at large distances from the obstacle the shock front forms a parabolic solid of revolution. The interior of the shock front is obtained by solution of the hydrodynamic equations in parabolic coordinates. The results are verified with a hydrodynamic simulation. The drag force and expected spectra are calculated for such shock, both in case of an optically thin and thick media. Finally, relations to astrophysical bow shocks and other analytic works on oblique shocks are discussed.
\end{abstract}

\section{Introduction}

Bow shocks occur when a supersonic flow encounters an obstacle.  Prominent examples from astrophysics include planetary bow shocks \cite{2008arXiv0808.1701T} and inundation of a dense  molecular cloud by a supernova shock wave \cite{McKee_1975}. A bow shock has even been observed around a star that moves at a super sonic velocity relative to the ISM \cite{Noriega_Crespo_1997}.

Bow shocks have been studied extensively, both theoretically \cite{Wilkin_1996,Farris_1994} and numerically \cite{Mohamed_2012, Miceli_2006}. However, all studies were carried out under the assumption of a finite Mach number. In this work we rather assume that the incoming matter is cold, so for every finite velocity its Mach number would be infinite. The asymptotic shape far from the obstacle is qualitatively different between the two. In the case of a finite Mach number $M$, far away from the obstacle the length scale of the obstacle becomes irrelevant, and the shock front coincides with the Mach cone, i.e. a cone with an opening angle $ \alpha = \sin^{-1} \frac{1}{M} $  \cite{LANDAU_1987}. When the Mach number tends to infinity, the opening angle of the Mach cone tends to zero. Hence the bow shock in that case cannot be straight, and the curvature must still be significant at large distances from the obstacle. In this paper we show that the shape of the bow shock front is in fact a parabola.

This paper is organized as follows. In section \ref{sec:math_form} we describe the complete mathematical formulation. In section \ref{sec:numerical_simulations} we present validation of our analytic results with a numerical simulation. In section \ref{sec:corollaries} we calculate the drag force of such shock. In section \ref{sec:application} we present a manifestation of this solution in an observation. Finally, in section \ref{sec:discussion}, we discuss the results.

\section{Mathematical Formulation} \label{sec:math_form}

\subsection{Steady State}

Spherical parabolic coordinates are given by

\begin{equation} x = \sigma \tau \cos \phi \end{equation}

\begin{equation} y = \sigma \tau \sin \phi \end{equation}

\begin{equation} z = \frac{1}{2} \left( \tau^2 - \sigma^2 \right) \end{equation}

In these coordinates, the steady state, azimuthally symmetric hydrodynamics equations take the following form. The conservation of mass is

\begin{equation} \frac{\partial}{\partial \sigma} \left( \rho \sqrt{\sigma^2 +\tau^2} \sigma \tau v_{\sigma} \right) + \frac{\partial}{\partial \tau} \left( \rho \sqrt{\tau^2 + \sigma^2} \sigma \tau v_{\tau} \right) = 0 \end{equation}

The conservation of entropy is 

\begin{equation}
v_\sigma \frac{\partial s}{\partial \sigma} + v_{\tau} \frac{\partial s}{\partial \tau} = 0
\end{equation}

where $ s = \ln p - \gamma \ln \rho $ is the specific entropy.
The conservation of momentum in each direction is

\begin{equation}
v_{\sigma} \frac{\partial v_{\sigma}}{\partial \sigma} + v_{\tau} \frac{\partial v_{\sigma}}{\partial \tau} + \frac{v_{\tau} \left(v_{\sigma} \tau - v_{\tau} \sigma \right)}{\tau^2+\sigma^2} + \frac{1}{\rho} \frac{\partial p}{\partial \sigma} = 0
\end{equation}

\begin{equation}
v_{\sigma} \frac{\partial v_{\tau}}{\partial \sigma} + v_{\tau} \frac{\partial v_{\tau}}{\partial \tau} - \frac{v_{\sigma} \left(v_{\sigma} \tau - v_{\tau} \sigma \right)}{\tau^2+\sigma^2} + \frac{1}{\rho} \frac{\partial p}{\partial \tau} = 0
\end{equation}

One of the momentum equations can be replaced by Bernoulli's equation

\begin{equation}
\frac{1}{2} v_{in}^2 = \frac{1}{2} v_{\sigma}^2 + \frac{1}{2} v_{\tau}^2 + \frac{1}{\gamma-1} c^2
\end{equation}
where $c$ is the speed of sound. We found it most convenient to express the equations in terms of the density $\rho$, the sound speed $c$, and the two Mach numbers $m_{\sigma,\tau} = \frac{v_{\sigma,\tau}}{c} $

We assume that the shock front coincides with a curve on which $\sigma = \sigma_s $ is constant. The Rankine Hugoniot boundary conditions at the shock fronts are

\begin{equation}
\rho_s = \frac{\gamma+1}{\gamma-1} \rho_a
\end{equation}

\begin{equation}
c_s = \frac{\sqrt{2\gamma \left( \gamma - 1 \right)}}{\gamma + 1} v_{in} \frac{\sigma_s}{\sqrt{\tau^2 + \sigma_s^2}}
\end{equation}

\begin{equation}
m_{\sigma s} = - \sqrt{\frac{\gamma - 1}{2 \gamma}}
\end{equation}

\begin{equation}
m_{\tau s} = \frac{\gamma + 1}{\sqrt{2 \gamma \left( \gamma - 1 \right) } } \frac{\tau}{\sigma_s}
\end{equation}
At very high values of $ \tau \gg \sigma_s $, the equations can be simplified. We assume that the variables vary with $ \tau $ as the shocked values do, i.e. $ \rho $ is independent of $ \tau $, $ c \propto \frac{1}{\tau} $, $ m_{\sigma} $ is independent of $\tau$ and $m_{\tau} \propto \tau $. Using this approximation, $\tau$ can be eliminated from the hydrodynamic equations, and the problem reduces to a set of ordinary differential equation in $\sigma$. These equation can be numerically integrated, and we can obtain curves for $\rho$, $\tau \cdot c$, $m_{\sigma}$ and $\frac{m_{\tau}}{\tau}$. The asymptotic equations are
\begin{equation}
2 m_{\sigma} + 2 \left( m_{\tau} \frac{\sigma}{\tau} \right) + m_{\sigma} \frac{d \ln \rho}{d \ln \sigma} + m_{\sigma} \frac{d}{d\ln \sigma} \ln \left( c \frac{\tau}{\sigma} \right) = 0
\end{equation}
\begin{equation}
2 m_{\sigma} - 2 \left( m_{\tau} \frac{\sigma}{\tau} \right) - \left( \gamma - 1 \right) m_{\sigma} \frac{d \ln \rho}{d \ln \sigma} + 2 m_{\sigma} \frac{d}{d \ln \sigma} \ln \left( c \frac{\tau}{\sigma} \right) = 0 
\end{equation}
\begin{equation}
-m_{\sigma}^2 - \frac{2}{\gamma} + m_{\sigma} \frac{d m_{\sigma}}{d \ln \sigma} + \frac{1}{\gamma} \frac{d \ln \rho}{d \ln \sigma} + \left( m_{\sigma}^2 + \frac{2}{\gamma} \right) \frac{d}{d \ln \sigma} \ln \left( c \frac{\tau}{\sigma} \right) = 0
\end{equation}
\begin{equation}
-m_{\sigma}^2+\frac{2}{\gamma}+m_{\sigma} \frac{d}{d \ln \sigma} \left( m_{\tau} \frac{\sigma}{\tau} \right) + m_{\sigma} \left( m_{\tau} \frac{\sigma}{\tau} \right) \frac{d}{d \ln \sigma} \left( c \frac{\tau}{\sigma} \right) = 0 
\end{equation}

\begin{figure}[ht!]
\begin{center}
\includegraphics[width=0.7\columnwidth]{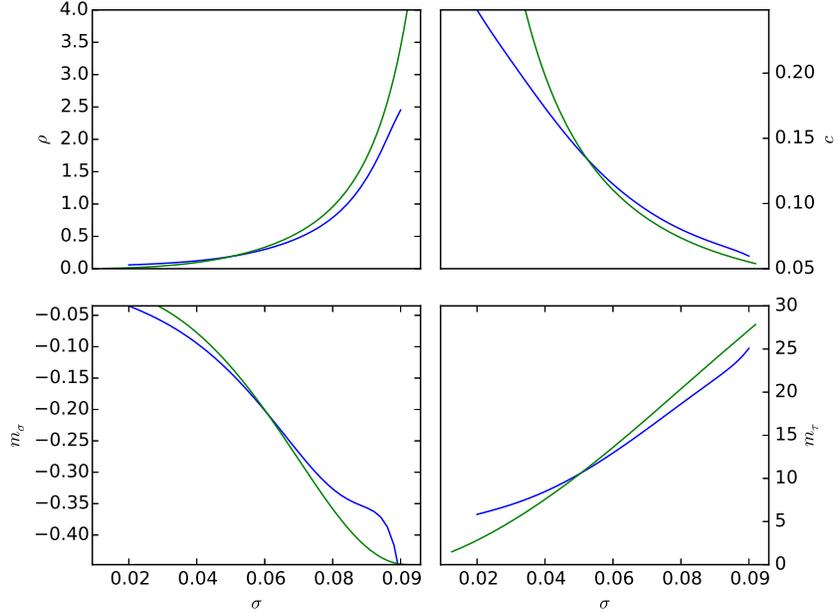}
\caption{Comparison of numerical result (blue) with analytic result (green) at $\tau = 1.5$. Top left is density, top right is the speed of sound, bottom left is $\sigma$ Mach number and bottom right is $\tau$ Mach number. \label{profiles}
}
\end{center}
\end{figure}

The complete analysis can be found in the following \href{https://www.dropbox.com/s/dufutv8t1h34i1o/spherical_parabolic_shock.nb?dl=0}{mathematica notebook}. These equations can be numerically integrated from the shock front to the symmetry axis to produce the complete profile. An example of such profile, for $\gamma = \frac{5}{3}$ can be seen in figure \ref{profiles}, along with a comparison to a numerical simulation, which will be described in the next section.

\subsection{Growth of the Shocked Region}
We describe the shock in a cylindrical coordinate system, where the $\hat{z}$ axis coincides with the symmetry axis. In this section we consider a shocked region of finite extent, so there is some point where the $r$ coordinates attains a maximum, and we denote that maximum by  $\varpi$. The rate at which wind adds mass to the shocked region is $ \pi \rho_a v_{in} \varpi^2$. Matter can only be compressed by a factor of a few (for reasonable values of $\gamma$), so the volume of the bow shock $V \propto \varpi^2 z \propto \varpi^4$ has to grow at the same rate. 

\begin{equation}
\frac{d}{dt} \left( \varpi^4 \right) \propto \varpi^2 \Rightarrow \varpi \propto t^{1/2}
\end{equation}

Therefore, the farthest point from the obstacle on the symmetry axis (on the lee side) moves at a constant velocity $z\left( \varpi \right) \propto \varpi^2 \propto t$. In the absence of other scales, that velocity has to be of the same order of magnitude as the incident velocity. 
For shocks with a finite Mach number $z \left( \varpi \right) \approx \min \left( v_{in} t, R \left( M^2 - 1\right) \right) $. Thus the age of a bow shock can be inferred from its size.

\begin{figure}[ht!]
\begin{center}
\includegraphics[width=0.7\columnwidth]{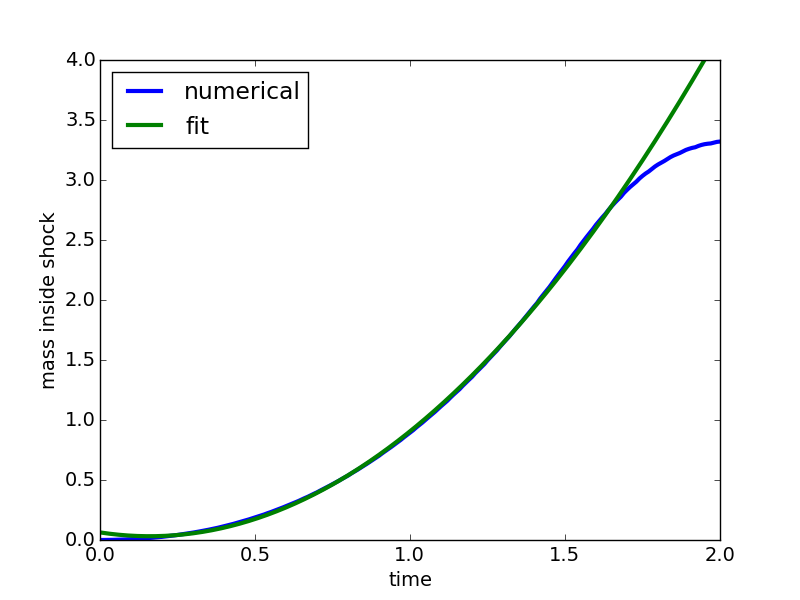}
\caption{History of the mass enclosed inside the bow shock. Comparison of numerical result (blue) with a parabolic fit (green). Deviation at late times is due to truncation of the bow shock by the boundaries of the computational domain. \label{fig:mass_history}
}
\end{center}
\end{figure}

These results were verified with a numerical simulation (see next section). We examined the history of the mass enclosed withing the bow shock. From the discussion above, it is clear that the mass scales with time as $M \propto V \propto \varpi^4 \propto t^2 $. Figure \ref{fig:mass_history} shows that indeed the mass is parabolic in time.

\section{Numerical Simulation} \label{sec:numerical_simulations}
\begin{figure}[ht!]
\begin{center}
\includegraphics[width=1.1\columnwidth]{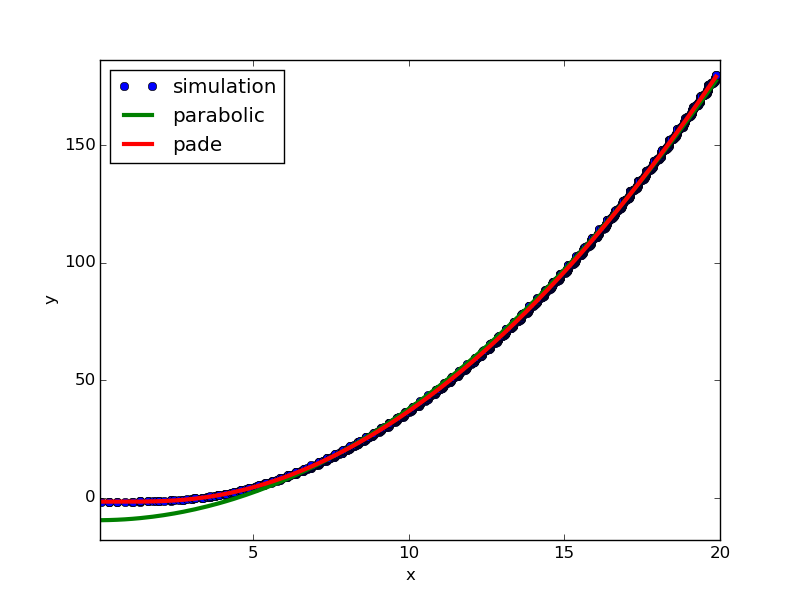}
\caption{Comparison between Pade fit (red) parabolic fit (green) and the shock front obtained from the simulation. \label{fig:pade}
}
\end{center}
\end{figure}

To verify our results, we ran a simulation in the RICH code\cite{Yalinewich_2015} (\href{https://github.com/bolverk/huji-rich/tree/version_2}{version 2 branch}). We have also released a complete listing of the \href{http://pastebin.com/8hxCdC6w}{source code} for the simulation. The simulations are in 2D cylindrical coordinates, assuming azimuthal symmetry. The boundaries of the computational domain are: $1.8>z>-0.2$ and $0.5>r>0$. The wind velocity is 1, in the positive $z$ direction. The obstacle is a circle of radius 0.01, located at the origin $\left(0,0\right)$. We ran the simulation to time $t=10$, and compared the numerical results to our analytic predictions.
First, we fit the shock front to an even parabola (figure \ref{fig:pade}), i.e. $y = a x^2 + b$. As can be seen from the figure, the shock front does converge to a parabola when the distance from the obstacle is much larger than the radius. The coefficients obtained from the simulation are $a = 0.474$ and $b = -9.67$. The ratio between the curvature radius of the shock front and the radius of the obstacle in this simulation is $\xi = \frac{R}{R_o} = 1.06$.
Next, we compared the numerical profiles inside the bow shock to the analytic prediction (figure \ref{profiles}). In order to do that, we interpolated the variables on the curve $\tau = 1.5$. The agreement between the two is reasonable. The deviations are due to two main reasons. The first is that since we are working in cylindrical coordinates, the relative error at each radius is equal to the ratio between the cell's width and the radius. The second is that the slope of the density is very steep next to the shock front, so a very fine resolution is required to resolve it.
In the previous section it was shown analytically that very far from the obstacle the shock is parabolic, but the analytic theory did not constrain the shape of the shock front close to the obstacle (henceforth referred to as the "nose"). Since the only length scale in this problem is the size of the obstacle, then the shape of the bow shock should scale as $ \frac{z}{R} = f \left( \frac{r}{R} \right)$. The dimensionless function $f$ can be obtained from the simulation. The ansatz chosen to represent $f$ is a six parameter Pade approximant in even powers of its argument
\begin{equation}
f \left( \psi \right) = \frac{c_0 + c_1 \psi^2 + c_2 \psi^4 + c_3 \psi^6}{1 + c_4 \psi^2 + c_5 \psi^4}
\label{eq:ansatz} \end{equation}
Odd powers were excluded to avoid a cusp at the $z$ axis. The choice of powers in the numerator and denominator the analytic asymptotic behavior $z \propto r^2$ for $r \rightarrow \infty$. The calibrated coefficients are given in table \ref{table:coefficients}. A comparison between the numerical is presented in figure \ref{fig:pade}, and the deviations between the two are shown to be quite small.

\begin{table}[ht]
\begin{tabular}{rc}
& Value \\
$c_0$ & -1.30 \\
$c_1$ & -0.29 \\
$c_2$ & -0.0548 \\
$c_3$ & 0.0215 \\
$c_4$ & 0.392 \\
$c_5$ & 0.0404 \\
\end{tabular}
\caption[]{Coefficients for the Ansatz \ref{eq:ansatz}. \\\hspace{\textwidth}Calibrated from the simulation.}
\label{table:coefficients}
\end{table}

\section{Drag Force} \label{sec:corollaries}

From dimensional analysis one can infer an expression for the drag force

\begin{equation}
F_d = c_d \pi R_0^2 \rho_a v_{in}^2 
\end{equation}
where $c_d$ is a dimensionless drag coefficient. It is possible to obtain an analytic upper bound on $c_d$. The highest pressure on the obstacle is at the stagnation point. This is the point on the surface of the obstacle that is closest to the nose of bow shock. The component of the velocity normal to obstacle vanishes on the surface of the obstacle, and due to symmetry considerations, the tangential velocity also vanishes at that point. Therefore, using Bernoulli's equation and the adiabatic relations the pressure can be evaluated there \cite{McKee_1975}

\begin{equation}
P_o = \frac{2}{\gamma + 1} \left( \frac{5 \gamma ^2 -2 \gamma + 1}{4 \gamma^2} \right)^{\frac{\gamma}{\gamma -1}} \rho_a v_{in}^2
\end{equation}
So the upper limit on the drag force is $ \pi R^2 P_o$, and the upper limit on the drag coefficient is $c_d = 0.827$ (for $\gamma = \frac{5}{3}$). 
The drag coefficient can be evaluated numerically, using the pressure on the obstacle $P \left( \theta \right)$, where $\theta$ is the angle measured relative to the direction of the wind. If the pressure is known, then the drag force is given by

\begin{equation}
F_d = \pi R^2 \int_0^{\pi} P \left( \theta \right) \sin \left( 2 \theta \right) d \theta
\end{equation}
For the simulation described above, the numerical drag coefficient is $c_d = 0.426$, which is approximately half of the upper limit discussed above.
The assumption of steady state can only be justified if the obstacle travels a distance much larger than its own radius before it experiences a significant change in velocity. This is equivalent to the condition that the density of the object be much larger than that of the ambient medium.

\section{Application to RX J1856.5-3754} \label{sec:application}
\begin{figure}[ht!]
\begin{center}
\includegraphics[width=1.1\columnwidth]{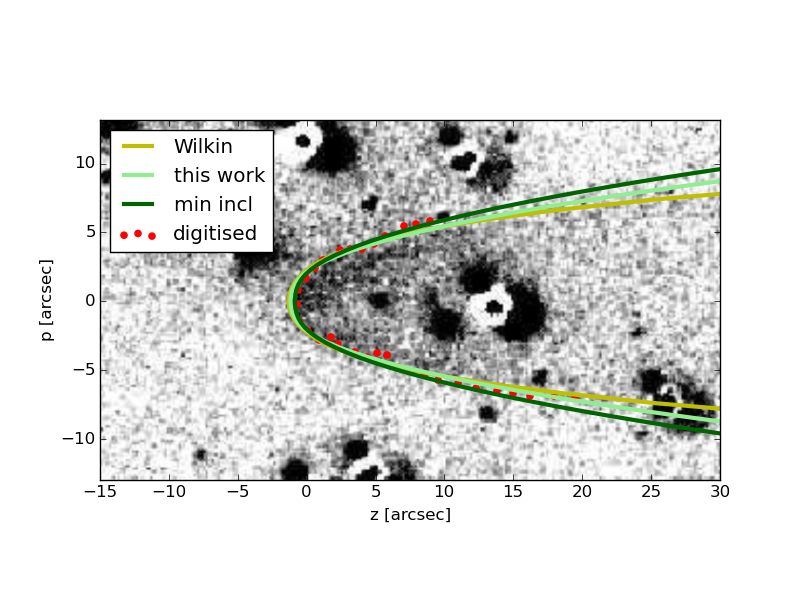}
\label{fig:observation}
\caption{Comparison of our model (light green) with Wilkin's model (yellow) on top of the original observation (gray - scale), taken from \cite{van_kerkwijk_kulkarni_2001}. Manually digitized points used for the fit are in red. Best fit with the inclination from \cite{van_kerkwijk_kulkarni_2001} is in light green. Fit with smallest inclination is in dark green. Our models seems to agree with observation. The fact that the nose is flatter then our model predicts is consistent with the simulation.}
\end{center}
\end{figure}

RX J1856.5-3754 is a neutron star that plows through the ISM at a speed of about $100 \frac{km}{s}$, and is adorned by a bow shock \cite{van_kerkwijk_kulkarni_2001}. The distance between the neutron star and the nose of the bow shock is about $1''$, and its tail extends to as far as $25''$. Its distance is about \href{http://chandra.harvard.edu/photo/2002/0211/}{400 light years}, and its inclination (angle between the line of sight and its velocity) is about $60^{\circ}$. The latter was inferred from measurements of proper motion and radial velocity.
A comparison between our model and the observations can be seen in figure \ref{fig:observation}. In order to perform the fit, we first manually digitized several points along the shock front. Van Kerkwijk and Kulkarni \cite{van_kerkwijk_kulkarni_2001} fit the observations to Wilkin's model \cite{Wilkin_1996}. The shock was shown to be adiabatic, whereas Wilkin's model assumes fast cooling. The figure below shows that our adiabatic model to the observation. The analytic form used for our fit is the Ansatz from \ref{eq:ansatz}. We note that in this fit we only have one free parameter, which is the size of the obstacle. We find that the best fit is obtained when the obstacle size is 0.9'', which agrees with the value in \cite{van_kerkwijk_kulkarni_2001} ($1\pm 0.2''$).
In principle, it is also possible to use the fit between our model and the observations to estimate the inclination. Unfortunately, the data does not provide very stringent constraints, and only allows us to set a lower limit on the angle at 40 degrees. From figure \ref{fig:observation} it is clear that inclination affects the wings of a bow shock more than it does the nose. 

\section{Discussion} \label{sec:discussion}
We discussed the problem of a cold wind blowing on a stationary obstacle. We derived the steady state, azimuthally symmetric hydrodynamics equations in cylindrical parabolic coordinates. We were able to use the approximation of very large $\tau$ to reduce these equations into a set of ordinary differential equations in $\sigma$. We numerically integrated these equations, and compared them to a full hydrodynamic simulation.

In the case of a large, but finite, Mach number, the bow shock would start out parabolic, but eventually straighten out and converge with the Mach cone. The transition will occur when the parabolic curve $z = \frac{1}{2} \frac{r^2}{R} $ intersects the Mach cone $z = \sqrt{M^2-1} r$
\begin{equation}
r_t \approx R_o \sqrt{M^2 - 1}
\end{equation}

\begin{equation}
z_t \approx R_o \left( M^2 - 1 \right)
\end{equation}
Therefore, the parabolic solution will only be valid in the region $R M^2 \gg z \gg R$.

Further work is required in order to apply this model to most astrophysical bow shocks. For example, planetary bow shocks require magneto - hydrodynamics, supernova - cloud interaction involve a time varying wind, and the complete description of the bow shock around Betelgeuse requires radiative cooling \cite{Mohamed_2012}. 

However, this model may apply to "knots" in supernova remnants \cite{fesen_2006}. These are bright, filamentary structures observed inside supernova remnants, but also sometimes outside the shock front. It is theorized that a fraction of the ejecta is in the form of small, dense gas clumps, and the filaments are the shocks they leave in their wake. Since they are much more compact than the rest of the ejecta, they are less affected by deceleration, and can therefore overtake the shock front. While inside the supernova remnant, their velocity is of the same order of magnitude as the bulk and thermal velocity of the ejecta. However, once they emerge from the shock front, the bow shock that will develop around them will be the epitome of the solution presented here.

Another utility of our solution is that it provides a lower bound on the width of an adiabatic bow shock. Bow shocks with lower mach numbers are bound to be wider than the solutions described here.

This research was partially supported by ISF and
iCore grants.

We would like to thank Marten van Kerkwijk for allowing us to use the image from his paper. AY would like to thank Elad Steinberg, Nir Mandelker and Prof. Chris McKee for enlightening discussions.

\bibliographystyle{apalike}
\bibliography{main.bib}

\end{document}